\documentclass[twocolumn,showpacs,preprintnumbers,amsmath,amssymb,superscriptaddress,prl]{revtex4}
\usepackage{epsfig}
\usepackage{graphicx}
\usepackage{color}
\usepackage{bm}

\begin{document}

\title{Electrical measurement of antiferromagnetic moments in exchange-coupled IrMn/NiFe stacks}
\author{X.~Mart\'{i}}
\affiliation{Faculty of Mathematics and Physics, Charles University in Prague, Ke Karlovu 3, 121 16 Prague 2, Czech Republic}
\affiliation{Institute of Physics ASCR, v.v.i., Cukrovarnick\'a 10, 162 53
Praha 6, Czech Republic}

\author{B.~G.~Park}
\affiliation{Hitachi Cambridge Laboratory, Cambridge CB3 0HE, United Kingdom}

\author{J.~Wunderlich}
\affiliation{Institute of Physics ASCR, v.v.i., Cukrovarnick\'a 10, 162 53
Praha 6, Czech Republic}
\affiliation{Hitachi Cambridge Laboratory, Cambridge CB3 0HE, United Kingdom}

\author{H.~Reichlov\'a}
\affiliation{Institute of Physics ASCR, v.v.i., Cukrovarnick\'a 10, 162 53
Praha 6, Czech Republic}
\affiliation{Faculty of Mathematics and Physics, Charles University in Prague, Ke Karlovu 3, 121 16 Prague 2, Czech Republic}
\author{Y.~Kurosaki}
\author{M.~Yamada}
\author{H.~Yamamoto}
\author{A.~Nishide}
\author{J.~Hayakawa}
\affiliation{Hitachi Ltd, Advanced Research Laboratory, 1-280 Higashi-koigakubo, Kokubunju-shi, Tokyo 185-8601, Japan}

\author{H. Takahashi}
\affiliation{Hitachi Cambridge Laboratory, Cambridge CB3 0HE, United Kingdom}
\affiliation{Hitachi Ltd, Advanced Research Laboratory, 1-280 Higashi-koigakubo, Kokubunju-shi, Tokyo 185-8601, Japan}

\author{T.~Jungwirth}
\affiliation{Institute of Physics ASCR, v.v.i., Cukrovarnick\'a 10, 162 53
Praha 6, Czech Republic} \affiliation{School of Physics and
Astronomy, University of Nottingham, Nottingham NG7 2RD, United Kingdom}

\begin{abstract}
We employ the recently discovered antiferromagnetic tunneling anisotropic magnetoresistance to study the behavior of antiferromagnetically ordered moments in IrMn exchange coupled to NiFe. Experiments performed by common laboratory tools for magnetization and electrical transport measurements allow us to directly link the broadening of the NiFe hysteresis loop and its shift (exchange bias) to the rotation and pinning of antiferromagnetic moments in IrMn.  At higher temperatures, the broadened loops show zero shift which correlates with the observation of fully rotating antiferromagnetic moments inside the IrMn film. The onset of exchange bias  at lower temperatures is linked to a partial rotation between distinct metastable states and  pinning of the IrMn antiferromagnetic moments in these states. The observation complements common pictures of exchange bias and reveals the presence of an electrically measurable memory effect in an antiferromagnet.
\end{abstract}


\maketitle

\begin{figure}[h]
\hspace*{0cm}\epsfig{width=.8\columnwidth,angle=0,file=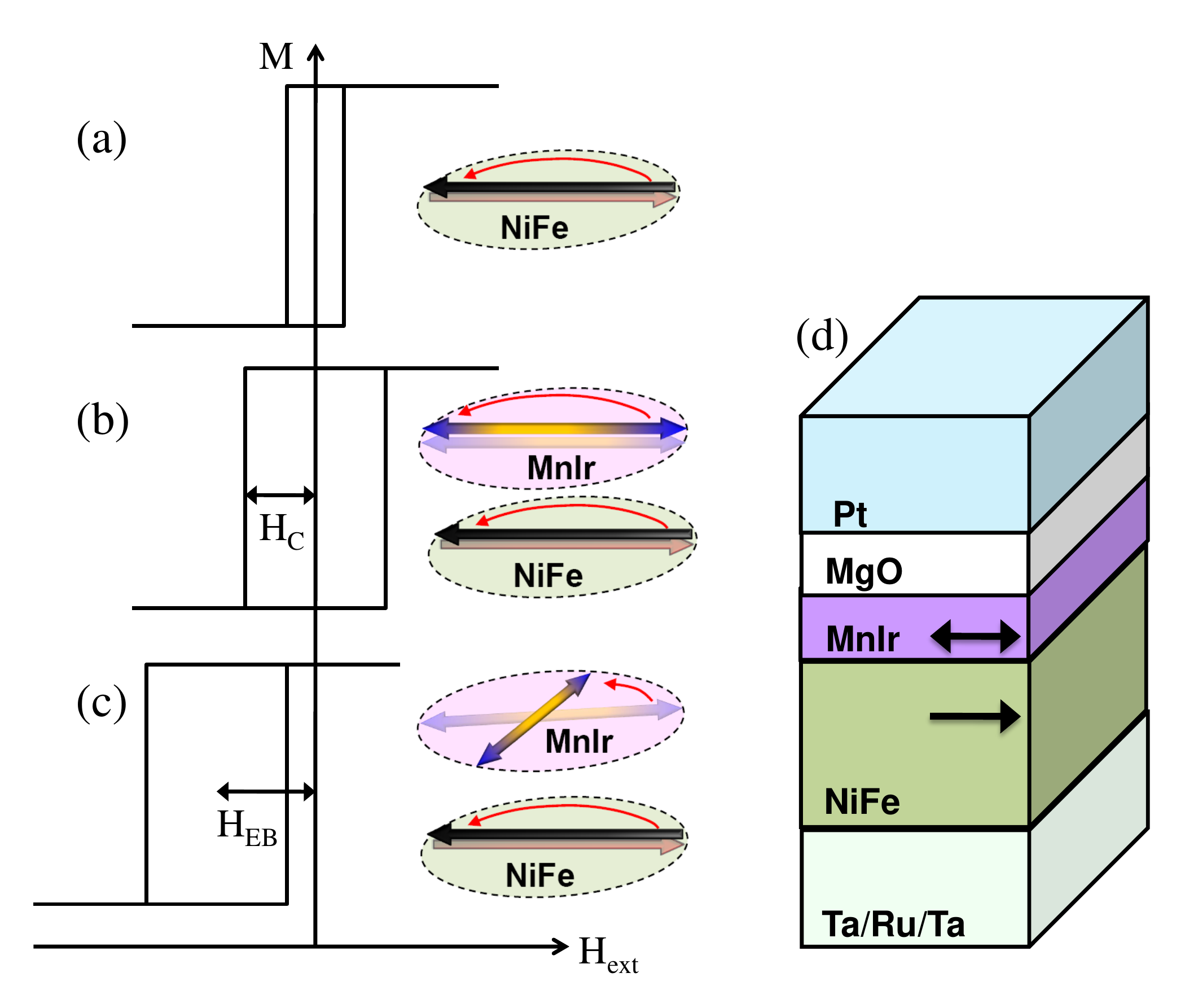}
\vspace*{-0.5cm}
\caption{Schematic illustration of the broadening ($H_c$) and shift ($H_{EB}$) of the hysteresis loop of a FM due to an exchange coupled AFM. (a) In a soft FM like NiFe, the hysteresis loop is narrow and centered around zero external magnetic field. (b) Broadening of the loop in a AFM/FM exchange coupled structure is ascribed to rotating AFM moments. (c) Exchange bias  occurs when the AFM moments remain pinned in their equilibrium position or tilt only partly during the magnetization reversal of the FM \protect\cite{Radu:2008_a}. (d) Schematic structure of the multilayers used in our experiments.}
\label{eb}
\end{figure}

Interlayer exchange coupling, giant magnetoresistance, and spin-transfer torque are the three key phenomena that have driven the development of current spintronic technologies \cite{Chappert:2007_a}. The oldest, yet least understood among the three effects is the exchange coupling, especially between an antiferromagnet (AFM) and a ferromagnet (FM) \cite{Radu:2008_a}. An AFM coupled to a FM can cause the broadening of the width, $H_c$, of the FM hysteresis loop, and the  shift, $H_{EB}$,  of the loop, as illustrated schematically in Figs.\ref{eb}(a)-(c). It is assumed that the former effect is due to the magnetic anisotropy of the AFM and the rotation of the AFM moments following the magnetization reversal in the coupled FM \cite{Radu:2008_a}. The exchange bias, on the other hand, has a character of an additional field which is ascribed to pinned (or only partially rotating) moments in the AFM \cite{Radu:2008_a}. 

A more detailed understanding of the AFM/FM exchange coupled structures requires a measurement of the orientation of AFM moments during the reversal of the FM. Studying AFMs is, however, notoriously difficult as compared to FMs. Moreover, the AFM films are typically only several nm to several 10's of nm thick in the common AFM/FM stacks which limits the applicability of established techniques like neutron diffraction. Synchrotron based X-ray linear dichroism experiments  on NiO/Co structures are among the very few measurements reported to date which have provided an experimental insight into the behavior of AFM moments during magnetization reversal of the coupled FM \cite{Scholl:2004_a}. In other systems, including the extensively explored stacks with the IrMn exchange biasing AFM, the X-ray linear dichroism has not been reported and the possibility to perform measurements of the AFM films outside large scale experimental facilities has remained elusive independent of the employed AFM material.

In this paper we demonstrate that with common laboratory tools we are able to investigate the behavior of both the FM and the AFM during magnetization reversal in the archetypical IrMn/NiFe exchange coupled systems. Magnetization in the FM NiFe is measured by the superconducting quantum interference device (SQUID). AFM moments in IrMn are detected electrically using the recently discovered AFM tunneling anisotropic magnetoresistance effect \cite{Park:2010_a}. The high sensitivity of this electrical method allows us to observe the rotating AFM moments up to temperatures which are close to the N\'eel temperature of the AFM film. Our measurements of the AFM moments also reveal phenomena beyond the common models of exchange bias  which have considered a simplified uniaxial anisotropy character of the AFM.
\begin{figure}[h]
\hspace*{0cm}\epsfig{width=.8\columnwidth,angle=0,file=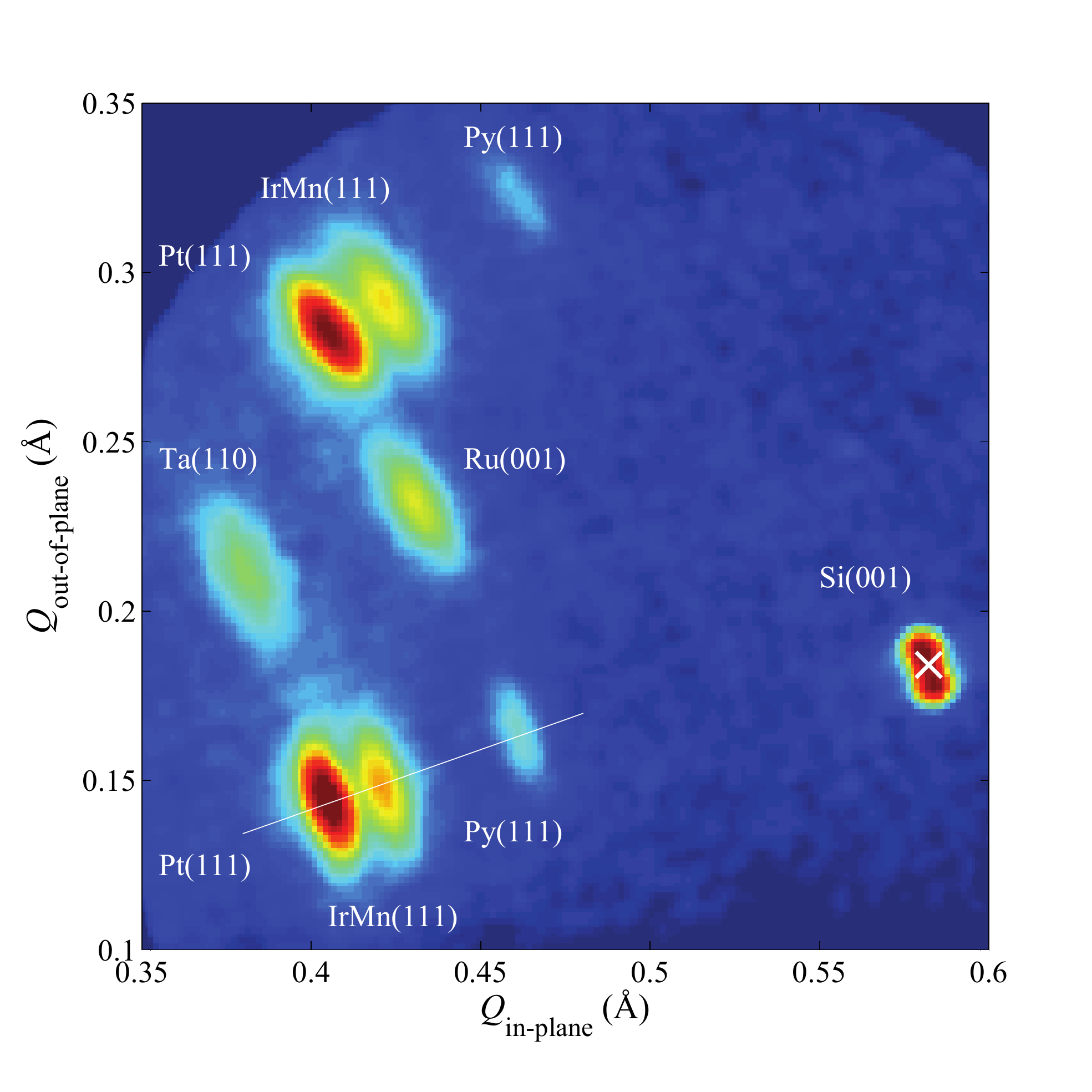}
\vspace*{-0.5cm}
\caption{X-ray diffraction reciprocal space map showing reflections from individual layers in the stack. From the location of reflections we conclude that the IrMn and NiFe layers are (111) out-of-plane textured. Crosses highlight the expected positions of the Si substrate. Line denotes the full-relaxation-line for cubic symmetry in the (111) reflection.
}
\label{xray}
\end{figure}

The multilayer structures used in our study, shown in Fig.~\ref{eb}(d), were deposited by the UHV RF magnetron sputtering on a  thermally oxidized 3" Si wafer substrate (700nm SiO$_2$ on (001) Si) with a base pressure of 10$^{-9}$~Torr. Multilayers of SiO$_2$/Ta(5)/Ru(10)/Ta(5)/Ni$_{0.8}$Fe$_{0.2}$(10)/Ir$_{0.2}$Mn$_{0.8}$(1.5, 3)/MgO(2.5)/Pt(10) were grown in a magnetic field of 5~mT along the flat edge direction of the wafer (layer thicknesses are given in nm). The MgO barrier was sputtered directly from a MgO target under a pure Ar atmosphere with the pressure of 10~mTorr.  Mesa structures of $1\times2\mu$m$^2$ - $5\times10\mu$m$^2$ were patterned from the wafer by photolithography and ion milling. After device fabrication, the wafer was annealed at 350$^{\circ}$C for 1h in a 10$^{-6}$~Torr vaccum in a magnetic field of 0.4~T applied along the same direction as during the growth. Several samples fabricated from each wafer were measured and showed comparable transport characteristics.  

The X-ray diffraction was used to verify the out-of-plane texture of the films. Data were acquired using a two-dimensional plate detector collecting the diffracted intensity while the sample is azimuthally rotating. Typical results for the 3~nm IrMn sample are shown in Fig~\ref{xray}. The peaks show the out-of-plane order in the stack. From their location it can be concluded that the IrMn and NiFe layers are (111) out-of-plane oriented (Ru and Ta are (001) and (110) oriented, respectively). The white cross in Fig~\ref{xray} showing the expected positions of the Si substrate indicate  the good alignment of the sample. The white line shows the predicted location of fully relaxed cubic-symmetry material in the (11-1) reflection. The overlap with the IrMn and NiFe indicates that the layers are fully relaxed.

\begin{figure}[h]
\hspace*{-0.2cm}\epsfig{width=1.07\columnwidth,angle=0,file=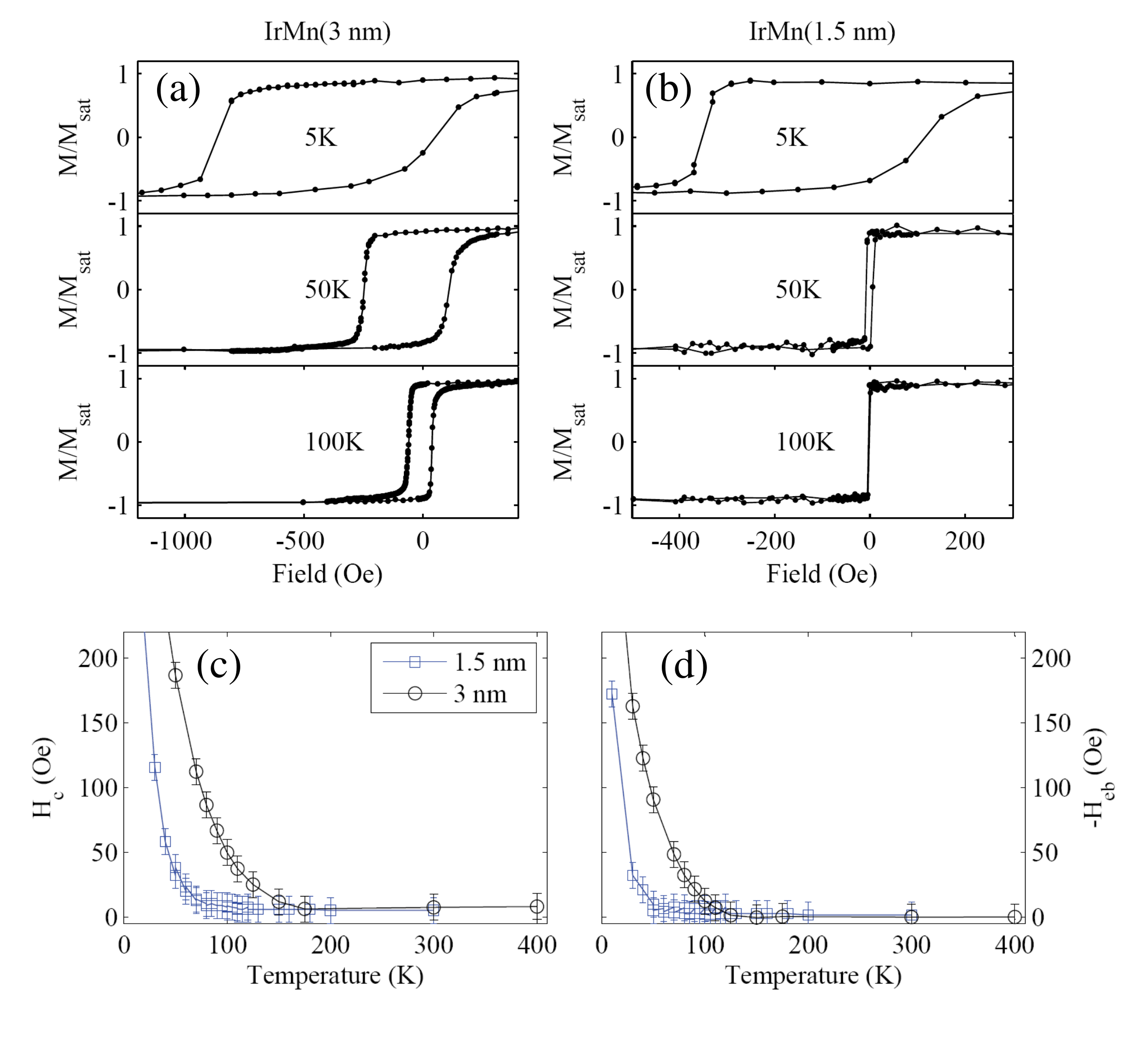}

\vspace*{-0.5cm}
\caption{SQUID magnetization loops of the 3~nm (a) and 1.5~nm (b) IrMn samples at 5, 50, and 100~K. (c), (d) Temperature dependence of the width of the hysteresis loop ($H_c$) and of the  shift of the loop ($H_{EB}$) for the two samples.
}
\label{squid}
\end{figure}

In Figs.~\ref{squid}(a),(b) we show magnetization loops measured by SQUID in the 3~nm and 1.5~nm IrMn samples at 5, 50, and 100~K. At 100~K, the 1.5~nm IrMn sample shows a very narrow hysteresis loop with coercive field $H_c\sim 10$~Oe which is similar to the coercivity of the reference NiFe sample without IrMn  and is comparable to the error bar of our SQUID measurements. At 50~K, $H_c$ of the 1.5~nm IrMn sample is enhanced but the hysteresis loop remains centered around zero external field, i.e., $H_{EB}$ is still zero at this temperature. Only at lower temperatures the broadening of the loop is accompanied by a non-zero  shift  as seen on the 5~K panel. The magnetization of the 3~nm  IrMn sample shows qualitatively the same behavior, only the onsets of the broadening of the hysteresis loop and of exchange bias occur at higher temperatures. This is illustrated in Fig.~\ref{squid}(b) and summarized in Figs.~\ref{squid}(c),(d).

\begin{figure}[h]
\hspace*{-.5cm}\epsfig{width=0.8\columnwidth,angle=0,file=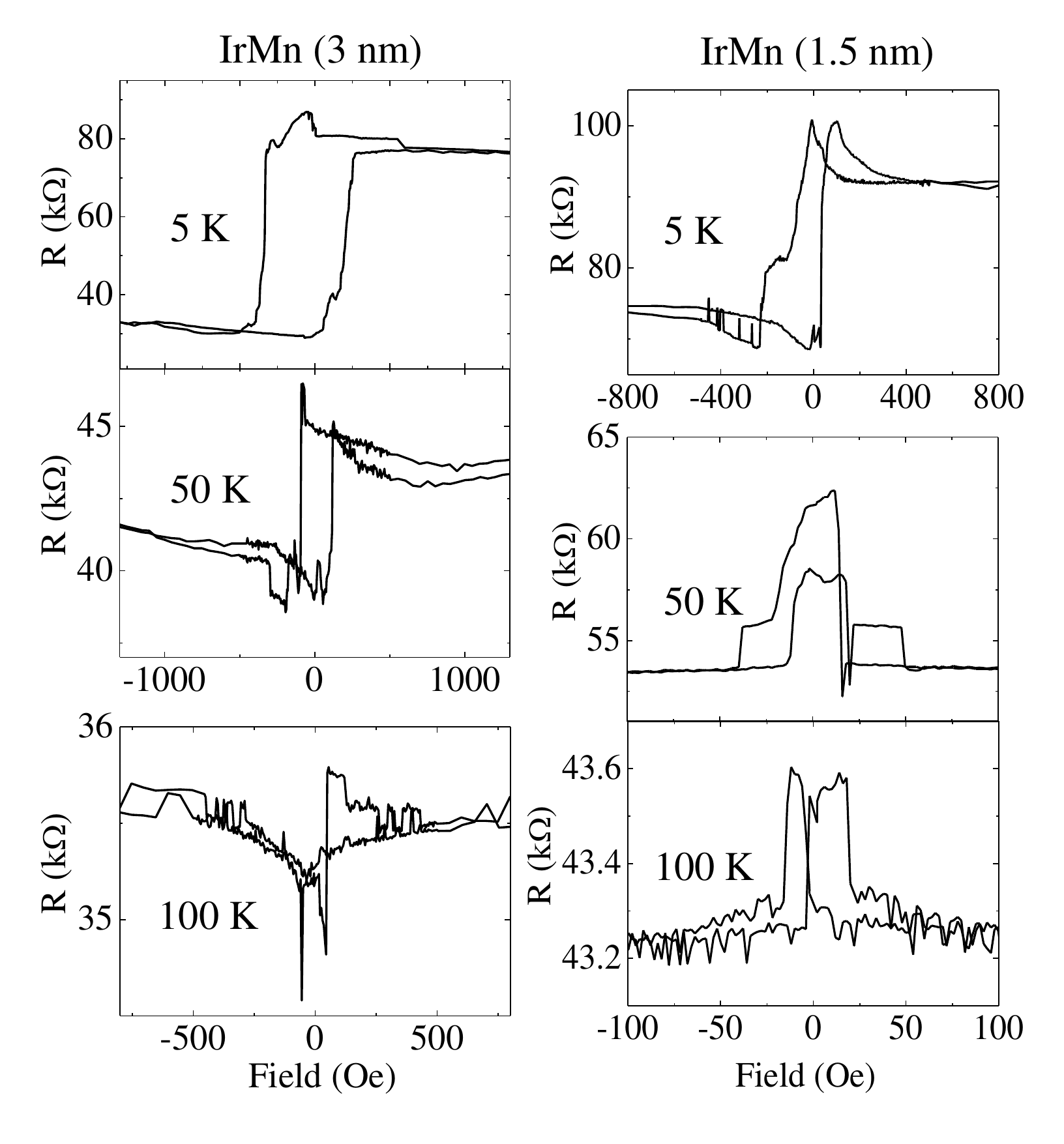}
\vspace*{-0.5cm}
\caption{Tunneling anisotropic magnetoresistance measured as a function of the applied magnetic field strength at the same conditions as the SQUID magnetization loops in Fig.~\protect\ref{squid}
}
\label{tamr}
\end{figure}

Results of our tunneling resistance measurements of the 3~nm and 1.5~nm IrMn samples at the same experimental conditions as in Fig.~\ref{squid} are shown in Fig.~\ref{tamr}. While the SQUID measurements detect the reversal of the FM moments, the tunneling magnetoresistance is sensitive to the change in the orientation of the AFM moments in IrMn. The origin of this transport signal is in the recently discovered tunneling anisotropic magnetoresistance of a AFM/insulator/normal-metal tunnel junction \cite{Park:2010_a}. In analogy to the magnetocrystalline anisotropy or optical linear dichroism, the tunneling anisotropic magnetoresistance is an even function of the microscopic moment and therefore  is present equally well for rotating AFM moments as for rotating moments in a FM \cite{Shick:2010_a,Park:2010_a}. Since the tunneling resistance is determined by layers adjacent to the tunnel barrier, in our device geometry (see Fig.~\ref{eb}(d)) it responds to the reorientation of the AFM moments in IrMn. The FM moments in the more remote NiFe layer have only an indirect effect on the tunneling transport; they can induce via an exchange spring effect \cite{Scholl:2004_a} the rotation of the AFM moments in IrMn during the magnetization reversal of the NiFe \cite{Park:2010_a}.

By comparing Figs.~\ref{squid} and \ref{tamr} we can draw a direct link between the behavior of FM moments in NiFe and AFM moments in IrMn during the reversal of NiFe. At 100~K, both the 3~nm and 1.5~nm IrMn samples do not show exchange bias in the SQUID measurements. Corresponding electrical transport mesurements show a resistance variation at low fields and the same resistance values at positive and negative saturation fields of the FM. This indicates that the AFM moments also undergo a full rotation, consistent with the schematic diagram in Fig.~\ref{eb}(b) for the broadened hysteresis loop and zero exchange bias. 

Note that in the 1.5~nm IrMn sample, the broadening of the SQUID hysteresis loops starts at temperatures close to 100~K but at 100~K it is still within the error of the magnetization measurement. The corresponding tunneling magnetoresistance signal is below 1\% at 100~K, however, it is clearly detectable. This illustrates the high sensitivity of our transport method for measuring the AFM moments in AFM/FM exchange coupled systems.

At 50~K, the 3~nm IrMn sample shows a qualitatively different behavior than the 1.5~nm IrMn sample. In the latter sample, the exchange bias is still zero and, consistently, the resistance values at negative and positive saturation fields of the FM are the same. The absence of the exchange bias is again linked with the full rotation of the AFM moments as was the case of the measurements at 100~K.   In the 3~nm IrMn sample, however, $H_{EB}$ measured by SQUID is non-zero at 50~K and the corresponding tunneling magnetoresistance trace shows different states of the AFM moments at large positive and negative fields. Consistent with the schematic diagram in Fig.~\ref{eb}(c), the AFM moments in IrMn tilt but do not make the complete rotation by 180$^{\circ}$ upon the full reversal of the FM moments in NiFe. The same correspondence between shifted magnetization loops of NiFe and asymmetric tunneling resistance traces of IrMn is observed at 5~K for both samples.

Our measurements are in broad agreement with the common picture of exchange bias as sketched in Fig.~\ref{eb}. We show this by recalling the Mieklejohn-Bean model \cite{Meiklejohn:1957_a,Radu:2008_a}.  As other common approaches it assumes an AFM with a uniaxial anisotropy. Exchange bias occurs in the model when $R\equiv K_{AF}t_{AF}/J_{eb}\ge1$, where $K_{AF}$ is the uniaxial anisotropy constant in the AFM, $t_{AF}$ is the thickness of the AFM, and $J_{eb}$ is the interfacial exchange energy. When the anisotropy energy of the AFM is decreased so that $R<1$ the exchange bias disappers. The FM magnetization loop can be broadened but shows zero shift in this regime. The behavior is explained in the model by the absence of the pinning of the AFM moments; above the blocking temperature, the AFM undergoes a complete 180$^\circ$ rotation upon the reversal of the FM. Our measurements in the 1.5~nm IrMn sample at 100~K and 50~K and in the 3~nm IrMn sample at 100~K are consistent with this picture when we assume that the ansiotropy energy of our thin IrMn films is suppressed at these higher temperatures. 

\begin{figure}[h]
\vspace*{0.5cm}
\hspace*{-0.6cm}\epsfig{width=0.8\columnwidth,angle=0,file=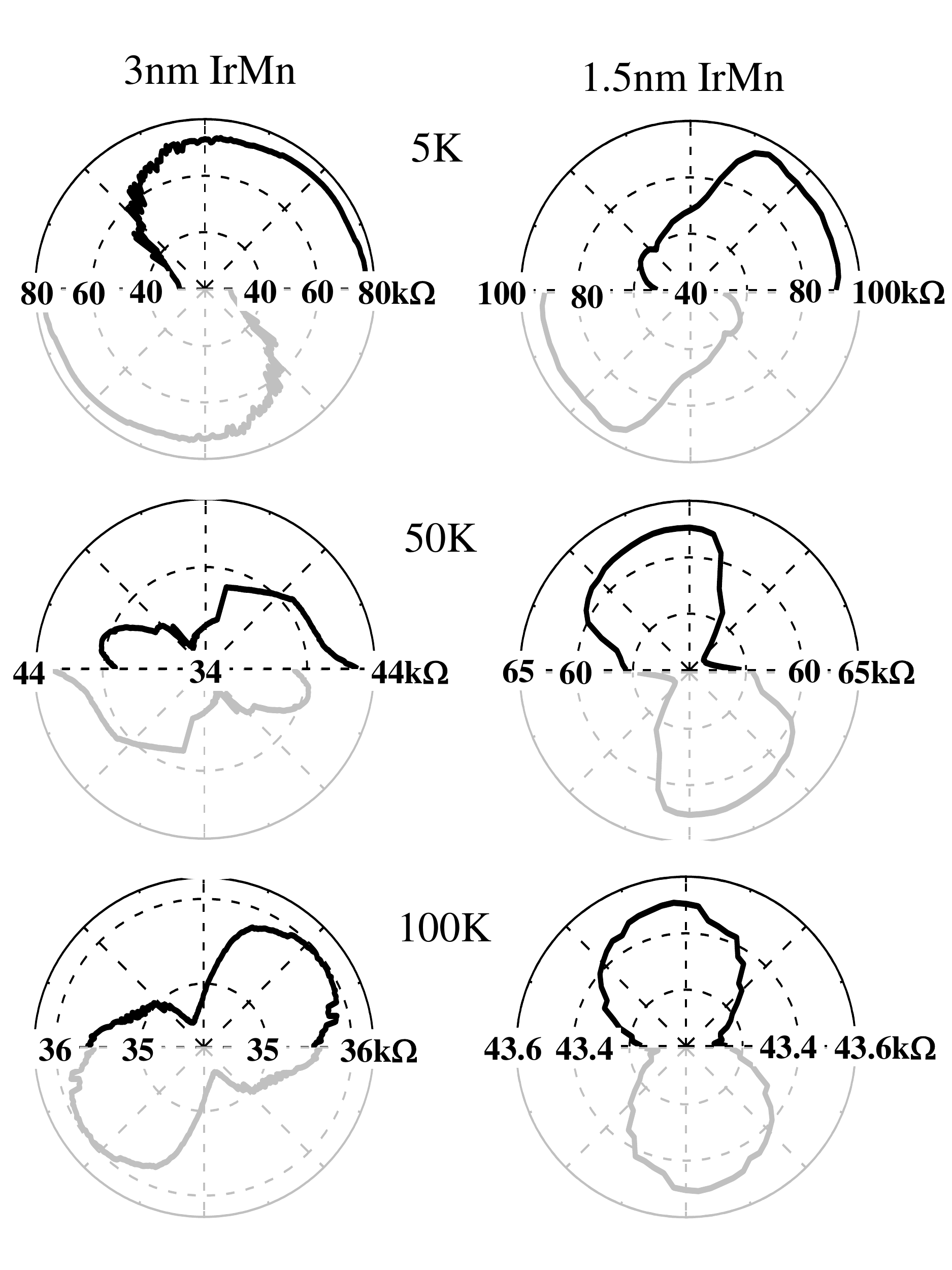}
\vspace*{-0.5cm}
\caption{Tunneling anisotropic magnetoresistance measured (black lines) as a function of the angle of the applied field with magnitude above the coercive field of the NiFe. Grey lines are inversion symmetry images of the measured data.
}
\label{tamr_rot}
\end{figure}

In the Mieklejohn-Bean model with a uniaxial AFM and for $R\ge1$, i.e. when exchange bias is non-zero, the AFM moments acquire only a small tilt (up to 45$^{\circ}$ for $R=1$) at intermediate external fields and return to the initial state when the 180$^\circ$ FM moment reversal is complete. The asymmetric magnetoresistance traces we observe at low temperatures indeed confirm that the AFM moments in IrMn remain partially pinned below the blocking temperature. However, in contrast with the model assumption of a uniaxial AFM, the data also demonstrate that the partial rotation of the AFM moments in IrMn occurs between  distinct metastable AFM configurations and the AFM moments do not rotate back to the initial state upon the full reversal of the NiFe FM moments. 

The bistability can be present at zero external magnetic field and the corresponding differences in the tunneling resistance can be $>100$\%, as seen e.g. in Fig.~\ref{tamr} in the panel for the 3~nm IrMn sample and 5~K. Our experiments, therefore, reveal a possibility for realizing memory effects in antiferromagnets which can be detected by large magnetoresistance signals. 

To further support our interpretation of the field-sweep tunneling anisotropic magnetoresistance traces shown in Fig.~\ref{tamr} we performed complementary transport measurements in which the amplitude of the applied field was fixed above the coercive field of NiFe and the magnetic field was rotated by 180$^\circ$. At this external field strength, the magnetization in NiFe follows the angle of the applied field. IrMn tunneling anisotropic magnetoresistance data recorded in the rotating field are plotted in Fig.~\ref{tamr_rot}. They are fully consistent with the picture inferred from the field-sweep measurements in Fig.~\ref{tamr}. For example at 5~K,  the resistance values are different at 0$^\circ$ and 180$^\circ$ and the resistance does not vary over a wide range of intermediate field angles. The AFM moments make a partial rotation and  then remain pinned. To highlight the lack of inversion symmetry in the data corresponding to partially pinned AFM moments we show in Fig.~\ref{tamr_rot} also the inversion symmetry images  of the measured field-rotation magnetoresistances. At 100~K, the resistance values are the same at 0$^\circ$ and 180$^\circ$ angles of the applied field. The AFM moments in IrMn make the full 180$^\circ$ rotation, i.e. follow the FM moments in NiFe. At 50~K, the full 180$^\circ$ rotation of the  AFM moments is observed only in the 1.5~nm IrMn sample, again consistent with the field-sweep measurements.

In previous studies of exchange-coupled AFM/FM systems, the onset of the broadening of $H_c$, followed by the onset of $H_{EB}$, has been shown to occur close to the N\'eel temperature of the AFM \cite{Hu:2004_b}. In our samples, the exchange bias, and the broadening of the FM hysteresis loop are undetectable above $\sim 150$~K in the 3~nm IrMn sample and above $\sim 100$~K in the 1.5~nm IrMn sample. The tunneling anisotropic magnetoresistance vanishes at similar temperatures as the broadening of $H_c$. The magnetization and transport data therefore consistently indicate that the N\'eel temperature decreases with decreasing thickness in the ultrathin IrMn films and is significantly suppressed compared to the bulk IrMn value. 

To conclude, we have employed a combined magnetization and electrical transport method to study FM and AFM moments in NiFe/IrMn exchange coupled systems. We have shown that these measurements provide an unprecedented experimental insight into the behavior of the IrMn AFM moments in the exchange bias phenomenon.  In future research we foresee the utility of the AFM anisotropic magnetoresistance in systematic thickness dependence studies of the ordering temperature and domains formation in the AFM. This will establish a more detailed understanding of the exchange bias effect. It will also guide research efforts towards room-temperature spintronic devices based on the AFM tunneling anisotropic magnetoresistance and other related spin-orbit coupling phenomena in AFMs.

We acknowledge support  from EU Grant FP7-214499 NAMASTE, ERC Advanced Grant 268066 0MSPIN, and from Czech Republic Grants from AV0Z10100521, LC510, and Preamium Academiae.


\begin{thebibliography}{7}
\expandafter\ifx\csname natexlab\endcsname\relax\def\natexlab#1{#1}\fi
\expandafter\ifx\csname bibnamefont\endcsname\relax
  \def\bibnamefont#1{#1}\fi
\expandafter\ifx\csname bibfnamefont\endcsname\relax
  \def\bibfnamefont#1{#1}\fi
\expandafter\ifx\csname citenamefont\endcsname\relax
  \def\citenamefont#1{#1}\fi
\expandafter\ifx\csname url\endcsname\relax
  \def\url#1{\texttt{#1}}\fi
\expandafter\ifx\csname urlprefix\endcsname\relax\def\urlprefix{URL }\fi
\providecommand{\bibinfo}[2]{#2}
\providecommand{\eprint}[2][]{\url{#2}}

\bibitem[{\citenamefont{Chappert et~al.}(2007)\citenamefont{Chappert, Fert, and
  Dau}}]{Chappert:2007_a}
\bibinfo{author}{\bibfnamefont{C.}~\bibnamefont{Chappert}},
  \bibinfo{author}{\bibfnamefont{A.}~\bibnamefont{Fert}}, \bibnamefont{and}
  \bibinfo{author}{\bibfnamefont{F.~N.~V.} \bibnamefont{Dau}},
  \bibinfo{journal}{Nature Mat.} \textbf{\bibinfo{volume}{6}},
  \bibinfo{pages}{813} (\bibinfo{year}{2007}).

\bibitem[{\citenamefont{Radu and Zabel}(2008)}]{Radu:2008_a}
\bibinfo{author}{\bibfnamefont{F.}~\bibnamefont{Radu}} \bibnamefont{and}
  \bibinfo{author}{\bibfnamefont{H.}~\bibnamefont{Zabel}},
  \bibinfo{journal}{Springer Tracts in Modern Phys}
  \textbf{\bibinfo{volume}{227}}, \bibinfo{pages}{97} (\bibinfo{year}{2008}).

\bibitem[{\citenamefont{Scholl et~al.}(2004)\citenamefont{Scholl, Liberati,
  Arenholz, Ohldag, and {St\"{o}hr}}}]{Scholl:2004_a}
\bibinfo{author}{\bibfnamefont{A.}~\bibnamefont{Scholl}},
  \bibinfo{author}{\bibfnamefont{M.}~\bibnamefont{Liberati}},
  \bibinfo{author}{\bibfnamefont{E.}~\bibnamefont{Arenholz}},
  \bibinfo{author}{\bibfnamefont{H.}~\bibnamefont{Ohldag}}, \bibnamefont{and}
  \bibinfo{author}{\bibfnamefont{J.}~\bibnamefont{{St\"{o}hr}}},
  \bibinfo{journal}{Phys. Rev. Lett.} \textbf{\bibinfo{volume}{92}},
  \bibinfo{pages}{247201} (\bibinfo{year}{2004}).

\bibitem[{\citenamefont{Park et~al.}(2011)\citenamefont{Park, Wunderlich,
  Marti, Holy, Kurosaki, Yamada, Yamamoto, Nishide, Hayakawa, Takahashi
  et~al.}}]{Park:2010_a}
\bibinfo{author}{\bibfnamefont{B.~G.} \bibnamefont{Park}},
  \bibinfo{author}{\bibfnamefont{J.}~\bibnamefont{Wunderlich}},
  \bibinfo{author}{\bibfnamefont{X.}~\bibnamefont{Marti}},
  \bibinfo{author}{\bibfnamefont{V.}~\bibnamefont{Holy}},
  \bibinfo{author}{\bibfnamefont{Y.}~\bibnamefont{Kurosaki}},
  \bibinfo{author}{\bibfnamefont{M.}~\bibnamefont{Yamada}},
  \bibinfo{author}{\bibfnamefont{H.}~\bibnamefont{Yamamoto}},
  \bibinfo{author}{\bibfnamefont{A.}~\bibnamefont{Nishide}},
  \bibinfo{author}{\bibfnamefont{J.}~\bibnamefont{Hayakawa}},
  \bibinfo{author}{\bibfnamefont{H.}~\bibnamefont{Takahashi}},
  \bibnamefont{et~al.}, \bibinfo{journal}{Nature Mat.}
  \textbf{\bibinfo{volume}{10}}, \bibinfo{pages}{347} (\bibinfo{year}{2011}).

\bibitem[{\citenamefont{Shick et~al.}(2010)\citenamefont{Shick, Khmelevskyi,
  Mryasov, Wunderlich, and Jungwirth}}]{Shick:2010_a}
\bibinfo{author}{\bibfnamefont{A.~B.} \bibnamefont{Shick}},
  \bibinfo{author}{\bibfnamefont{S.}~\bibnamefont{Khmelevskyi}},
  \bibinfo{author}{\bibfnamefont{O.~N.} \bibnamefont{Mryasov}},
  \bibinfo{author}{\bibfnamefont{J.}~\bibnamefont{Wunderlich}},
  \bibnamefont{and}
  \bibinfo{author}{\bibfnamefont{T.}~\bibnamefont{Jungwirth}},
  \bibinfo{journal}{Phys. Rev. B} \textbf{\bibinfo{volume}{81}},
  \bibinfo{pages}{212409} (\bibinfo{year}{2010}).

\bibitem[{\citenamefont{Meiklejohn and Bean}(1957)}]{Meiklejohn:1957_a}
\bibinfo{author}{\bibfnamefont{W.~H.} \bibnamefont{Meiklejohn}}
  \bibnamefont{and} \bibinfo{author}{\bibfnamefont{C.~P.} \bibnamefont{Bean}},
  \bibinfo{journal}{Phys. Rev.} \textbf{\bibinfo{volume}{105}},
  \bibinfo{pages}{904} (\bibinfo{year}{1957}).

\bibitem[{\citenamefont{guo Hu et~al.}(2004)\citenamefont{guo Hu, Jin, Hu, and
  qiang Ma}}]{Hu:2004_b}
\bibinfo{author}{\bibfnamefont{J.}~\bibnamefont{guo Hu}},
  \bibinfo{author}{\bibfnamefont{G.}~\bibnamefont{Jin}},
  \bibinfo{author}{\bibfnamefont{A.}~\bibnamefont{Hu}}, \bibnamefont{and}
  \bibinfo{author}{\bibfnamefont{Y.}~\bibnamefont{qiang Ma}},
  \bibinfo{journal}{Eur. Phys. J. B} \textbf{\bibinfo{volume}{40}},
  \bibinfo{pages}{265} (\bibinfo{year}{2004}).

\end{thebibliography}

\end{document}